\documentclass{emulateapj}

\newcommand{\sqcm}{${\rm cm^{-2}}$}

\newcommand{\mum}{${\rm \mu m}$}

\slugcomment{Submitted 26 March 2004; accepted 14 May 2004; scheduled
for publication in ApJS Sept. 2004, v154 [Spitzer Special issue]}

\shorttitle{Ices toward Low Mass Protostars}
\shortauthors{Boogert et al.}

\begin{document}

\title{Spitzer Space Telescope Spectroscopy of Ices toward Low Mass
Embedded Protostars\footnote{\tiny Some of the data presented herein
were obtained at the W.M. Keck Observatory, which is operated as a
scientific partnership among the California Institute of Technology,
the University of California and the National Aeronautics and Space
Administration. The Observatory was made possible by the generous
financial support of the W.M. Keck Foundation.}\footnote{\tiny The
VLT/ISAAC spectra were obtained at the European Southern Observatory,
Paranal, Chile, within the observing program 272.C-5008}}

\author{A. C. Adwin Boogert\altaffilmark{1}, 
        Klaus M. Pontoppidan\altaffilmark{2}, 
        Fred Lahuis\altaffilmark{2,3}, 
        Jes K. J{\o}rgensen\altaffilmark{2},
        Jean-Charles Augereau\altaffilmark{2}, 
        Geoffrey A. Blake\altaffilmark{4}, 
        Timothy Y. Brooke\altaffilmark{1}, 
        Joanna Brown\altaffilmark{1}, 
        C. P. Dullemond\altaffilmark{5}, 
        Neal J. Evans, II\altaffilmark{6},
        Vincent Geers\altaffilmark{2}, 
        Michiel R. Hogerheijde\altaffilmark{2}, 
        Jacqueline Kessler-Silacci\altaffilmark{6}, 
        Claudia Knez\altaffilmark{6},
        Pat Morris\altaffilmark{7}, 
        Alberto Noriega-Crespo\altaffilmark{7}, 
        Fredrik L. Sch{\"o}ier\altaffilmark{2},  
        Ewine F. van Dishoeck\altaffilmark{2},
        Lori E. Allen\altaffilmark{8}, 
        Paul M. Harvey\altaffilmark{6}, 
        David W. Koerner\altaffilmark{9}, 
        Lee G. Mundy\altaffilmark{10}, 
        Philip C. Myers\altaffilmark{8}, 
        Deborah L. Padgett\altaffilmark{7}, 
        Anneila I. Sargent\altaffilmark{1}, 
        Karl R. Stapelfeldt\altaffilmark{11}}

\altaffiltext{1}{Division of PMA, Mail Code 105-24, California
Institute of Technology, Pasadena, CA 91125, USA}
\altaffiltext{2}{Leiden Observatory, PO Box 9513, 2300 RA Leiden, the
Netherlands} 
\altaffiltext{3}{SRON, PO Box 800, 9700 AV Groningen, the Netherlands}
\altaffiltext{4}{Division of GPS, Mail Code 150-21, California
Institute of Technology, Pasadena, CA 91125, USA}
\altaffiltext{5}{Max-Planck-Institut f{\"u}r Astrophysik, P.O. Box 1317,
D-85741 Garching, Germany}
\altaffiltext{6}{Department of Astronomy, University of Texas at
Austin, 1 University Station C1400, Austin, TX 78712-0259}
\altaffiltext{7}{Spitzer Science Center, California Institute of
Technology, CA 91125, USA}
\altaffiltext{8}{Smithsonian Astrophysical Observatory, 60 Garden
Street, MS42, Cambridge, MA 02138}
\altaffiltext{9}{Department of Physics and Astronomy, Northern Arizona
University, Box 6010, Flagstaff, AZ 86011-6010}
\altaffiltext{10}{Department of Astronomy, University of Maryland,
College Park, MD 20742}
\altaffiltext{11}{Jet Propulsion Laboratory, MS 183-900, California
Institute of Technology, 4800 Oak Grove Drive, Pasadena, CA 91109}

\begin{abstract}
Sensitive 5-38 \mum\ Spitzer Space Telescope (SST) and ground based
3-5 \mum\ spectra of the embedded low mass protostars B5 IRS1 and HH46
IRS show deep ice absorption bands superposed on steeply rising
mid-infrared continua. The ices likely originate in the circumstellar
envelopes. The CO$_2$ bending mode at 15 \mum\ is a particularly
powerful tracer of the ice composition and processing history.  Toward
these protostars, this band shows little evidence for thermal
processing at temperatures above 50 K.  Signatures of lower
temperature processing are present in the CO and OCN$^-$ bands,
however. The observed CO$_2$ profile indicates an intimate mixture
with H$_2$O, but not necessarily with CH$_3$OH, in contrast to some
high mass protostars. This is consistent with the low CH$_3$OH
abundance derived from the ground based L band spectra. The
CO$_2$/H$_2$O column density ratios are high in both B5 IRS1 and HH46
IRS ($\sim 35$\%).  Clearly, the SST spectra are essential to study
ice evolution in low mass protostellar environments, and to eventually
determine the relation between interstellar and solar system ices.
\end{abstract}


\section{Introduction}~\label{sec:intro}

A recurring question in disk, planet and comet formation studies is
how the composition of molecular material evolves as it flows from the
molecular cloud to the protostellar envelope and protoplanetary disk.
Much of the material in these environments is frozen on grains.  A
plethora of processes, including heating by stellar photons, shocks
related to accretion or outflow, cosmic ray hits, and ultraviolet
irradiation can change the ice structure and composition.  The
spectroscopic effects of these processes can be observed in the
vibrational bands of the ices through infrared spectroscopy.  Ices
around high mass protostars have been extensively studied this
way. Infrared Space Observatory (ISO) spectra have shown that in
particular the ice structure is affected by heating from the central
star. The simplicity of the ice composition does indicate that the
formation of new species through ultraviolet irradiation or cosmic ray
hits occurs at a low level at best. Observations of ices around low
mass protostars have been limited due to the unavailability of much of
the 5-20 \mum\ spectral region, where many of the molecular bending
mode transitions occur. In particular the CO$_2$ bending mode at 15
\mum\ is a valuable tracer of ice structure and composition
(Ehrenfreund et al. 1998; Gerakines et al. 1999). With the sensitive
Infrared Spectrometer (IRS; \citealt{hou04}) on board of the Spitzer
Space Telescope (SST; \citealt{wer04}) this band can now be observed
for the first time at high quality in low mass systems.

\begin{figure}[!t]
\includegraphics[angle=90, scale=0.42]{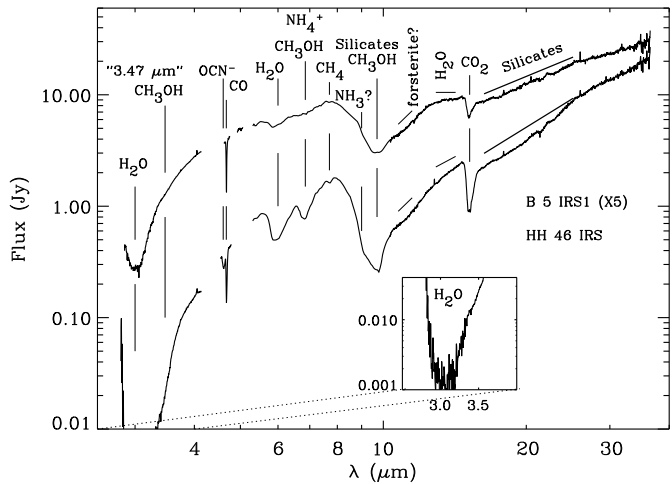}
\caption{Combined Spitzer Space Telescope and ground based L and M
band spectroscopy of B5 IRS1 (top, multiplied by factor of 5 for
clarity) and HH46 IRS (bottom). Identifications and possible
identifications are indicated.}~\label{f3}
\end{figure}

Observation of two protostars are presented in this paper.  B5 IRS1
(IRAS 03445+3242; Beichman et al. 1984; $L=$10 L$_\odot$) is well
studied at infrared and millimeter wavelengths (e.g. Charnley et
al. 1990; Langer et al. 1996).  The millimeter continuum emission is
resolved on a few arcsec scale, and may originate in an inclined
disk. The outflow has received most of the attention, and has a large
opening angle leading to significant outflow/infall interaction.  HH46
IRS (IRAS 08242-5050; $L=12\ $L$_\odot$) is also deeply embedded, and
is also the driving source of a powerful outflow. SST imaging and
spectroscopic observations of this source, focused on the outflow, are
presented in Noriega-Crespo et al. (2004).

\section{Observations}~\label{sec:obs}

B5 IRS1 and HH46 IRS were observed with SST/IRS as part of the `c2d'
Spitzer Legacy program (Evans et al. 2003) in the modules Short Low
(SL; $\lambda=5-14$ \mum; R=64-128), Short High (SH; $\lambda=10-20$
\mum; R=600) and Long High (LH; $\lambda=20-38$ \mum; R=600).  The
archival AOR keys are 0005638912 (HH46 IRS) and 0005635328 (B5 IRS1)
for PROGID 172. Both sources are well centered in all slits.  The
integration time was 28 seconds per module per source at 14 second
ramps, except for SH which has 24 seconds in total and 6 second
ramps. The spectra were reduced with the IRS pipeline version 9.5 on
10 March 2004 at the SSC. Bad pixels were interpolated in the spectral
domain in the 2D images before extracting 1D spectra.  Accurate
wavelength calibration was assured using calibration tables available
in May 2004. For overlapping SH spectral orders, the poorly calibrated
long wavelength part of each order was removed. This is particularly
important to obtain a reliable profile of the CO$_2$ ice band at 15.2
\mum, where two orders overlap.  HH46 IRS was observed independently
as an Early Release Observation (AOR key 0007130112 and PROGID 1063;
Noriega-Crespo et al. 2004). The datasets are in good agreement, and
were averaged. Finally, complementary ground based observations were
obtained. B5 IRS1 was observed with Keck/NIRSPEC \citep{mcl98} at
R=25000 in the M band and at R=2000 in the L band. VLT/ISAAC
\citep{moo99} L and M band spectra of HH46 IRS were obtained at R=600
and 5000 respectively. The SST and ground based spectra were combined
by scaling the ground based data.

\begin{figure*}[t]
\center
\includegraphics[angle=90, scale=0.80]{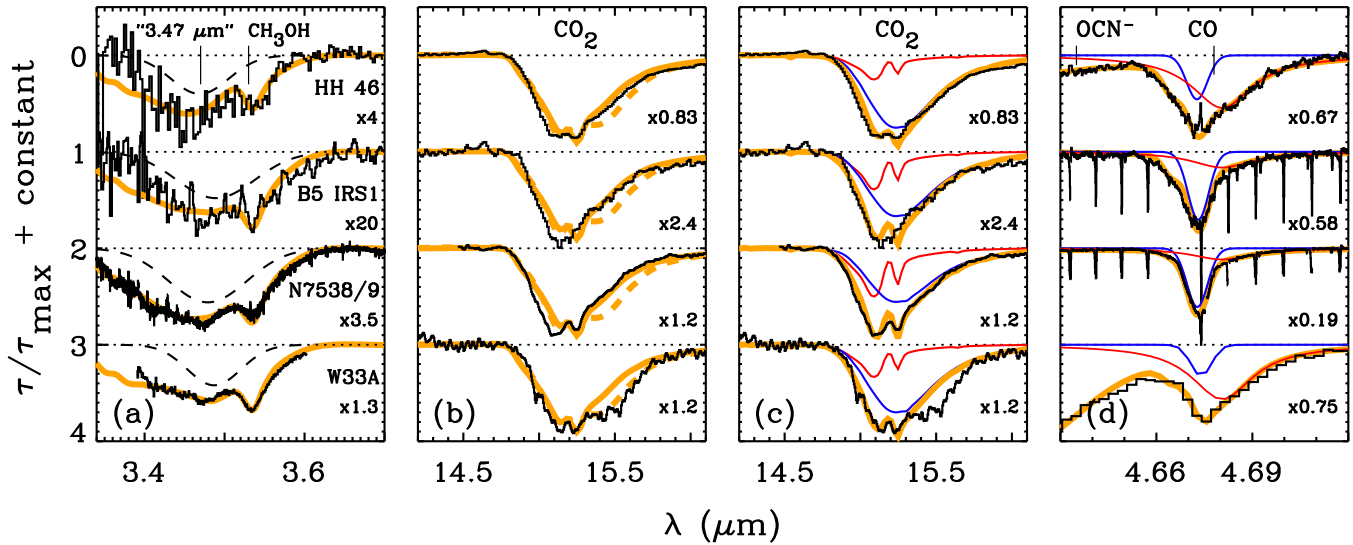}
\caption[]{Solid CH$_3$OH ({\bf a}), CO$_2$ ({\bf b and c}), and CO
({\bf d}) spectra of HH46 IRS and B5 IRS1 compared to the massive
protostars NGC7538 IRS9 and W33A in each panel from top to bottom.
The normalization factor is indicated in the right corner of each
plot.  {\bf Panel a:} a laboratory pure CH$_3$OH spectrum at $T_{\rm
lab}=10$ K (thick yellow line) to which is added a Gaussian (dashes)
to account for the underlying ``3.47 \mum'' absorption feature. The
NGC7538 IRS9 and W33A data are from Boogert et al. (2004) and Brooke
et al. (1999) respectively.  {\bf Panel b:} the CO$_2$ bending mode
observed with SST/IRS (HH46 IRS and B5 IRS1) and ISO/SWS (NGC 7538
IRS9 and W33A). The thick yellow dashed line is the laboratory ice
H$_2$O:CH$_3$OH:CO$_2$=0.84:0.92:1, and the continuous thick yellow
line is a laboratory ice with reduced CH$_3$OH content,
H$_2$O:CH$_3$OH:CO$_2$=0.92:0.29:1, both at $T_{\rm lab}$=115 K. Only
for W 33A an enhanced CH$_3$OH/CO$_2$ ratio is required, in agreement
with the strong 3.53 \mum\ CH$_3$OH feature in panel (a). {\bf Panel
c:} the CO$_2$ bending mode for the same sources as in panel (c), but
now fitted with different laboratory spectra: a highly processed polar
ice ($T_{\rm lab}$=125 K; red) in combination with an H$_2$O--rich,
CH$_3$OH-deficient cold ice ($T_{\rm lab}$=10 K; blue). The sum of
these two components is the thick yellow line. {\bf Panel d:} the
absorption band of solid CO observed with VLT/ISAAC (HH46 IRS),
Keck/NIRSPEC (B5 IRS1 and NGC7538 IRS9) and ISO/SWS (W33A).  Narrow
absorption lines are due to circumstellar gas phase CO. The onset of
the 4.62 \mum\ `XCN' band is seen in HH46 IRS and W33A. Gaussian and
Lorentzian fits to the apolar and polar CO components are indicated by
blue and red lines respectively. The thick yellow line represents the
combined fit, including the 'XCN' bands and blue-shifted CO component
that are not shown separately.}~\label{f2}
\end{figure*}

\section{Results}~\label{sec:res}

The combined SST and ground based 3--38 \mum\ spectra of B5 IRS1 and
HH46 IRS show steeply rising continua (Fig. 1). The continuum of HH46
IRS is the steepest, with a 35 \mum/4 \mum\ flux ratio of 100,
compared to 10 for B5 IRS1.  Numerous silicate and ice absorption
features are superposed. The main ice constituents are H$_2$O, CO$_2$,
and CO. The well known, yet unidentified, 3.47 and 6.85 \mum\ bands
are present in both sources as well.  Rarely seen before is a shallow
absorption feature at 11.2 \mum\ in both sources.  It may be related
to refractory dust components, such as crystalline forsterite
\citep{kes04}.  In addition, HH46 IRS, but not B5 IRS1, shows
absorption by CH$_4$ and `XCN' (likely OCN$^-$; van Broekhuizen et
al. 2004). Finally, other ice features may be present in the 5-10
\mum\ region (NH$_3$, HCOOH), but a dedicated analysis is required to
verify their reality. We focus on the band profiles and abundances of
the main ice components and their relation to the 15 \mum\ CO$_2$
bands, newly discovered with the SST.

The CO$_2$ bending mode has rarely been observed toward low mass
protostars, and never at such high quality. It is blended with the
short wavelength wing of the silicate bending mode. The CO$_2$ band is
put on an optical depth scale, assuming the `intrinsic' profile of the
silicate band is represented by the Galactic Center source GC3 (Chiar
et al. 2000). For this a third order polynomial is fitted to the
wavelength regions 13.0--14.7 and 26.3-33.3 \mum. The resulting CO$_2$
bands look similar to those observed in other lines of sight; note in
particular the presence of a long wavelength wing extending to at
least 16 \mum\ (Gerakines et al. 1999).  Similar to the massive
protostar NGC 7538 IRS9 (Fig. 2b), weak double ice crystallization
peaks are observed at the bottom of the band in HH46 IRS, but not in
B5 IRS1.  Neither B5 IRS1 nor HH46 IRS show evidence for a third peak
at 15.38 \mum, expected in CH$_3$OH:CO$_2$ complex formation.

To place the interpretation of the CO$_2$ band within a larger
perspective, ground based observations of the 3.53 \mum\ CH$_3$OH band
and the 4.67 \mum\ band of solid CO are analyzed. The 3.53 \mum\ band
is superposed on the wing of the strong 3.07 \mum\ H$_2$O band and is
locally blended with the unidentified 3.47 \mum\ band. Following
Brooke et al. (1999), we derive the continuum and separate the
CH$_3$OH contribution. For comparison, spectra of the massive
protostars W33A (Brooke et al. 1999) and NGC 7538 IRS9 (Boogert et
al. 2004) are analyzed as well.  Hints of CH$_3$OH are seen in both
HH46 IRS and B5 IRS1, resulting in column densities of 7\% with
respect to H$_2$O, comparable to NGC 7538 IRS9, but much less compared
to W33A. For HH46 IRS the detection of CH$_3$OH is strengthened by the
presence of a feature at 9.7 \mum\ in the bottom of the silicate band
(Fig. 1). Both HH46 IRS and B5 IRS1 show prominent bands of solid CO
at 4.67 \mum\ (Fig. 2d). The ratio between the central narrow CO
component and the broad long wavelength wing, representing the column
density ratio of volatile pure CO and CO mixed with less volatile
H$_2$O, is a factor of five smaller in HH46 IRS.  In fact, the profile
of HH46 IRS is comparable to that of the massive protostar W33A. The
latter two sources also show a band at 4.62 \mum, most likely
attributed to the OCN$^-$ ion. Relative to H$_2$O, the OCN$^-$ column
density is comparable to (upper limits to) those of B5 IRS1 and
NGC7538 IRS9.  Furthermore, the 4.67 \mum\ spectrum of B5 IRS1 shows
deep ro-vibrational gas phase CO lines. The presence of gas phase CO
in HH46 IRS is hard to assess because of telluric contamination. The
intriguing differences and similarities between HH46 IRS and B5 IRS1,
as well as compared to massive protostars, provide insight into the
formation and evolution of interstellar and circum-protostellar ices.

Finally, column densities of the main ices, summarized in Table 1, are
derived by dividing the integrated optical depth over the laboratory
integrated band strength (e.g. Hudgins et al. 1993). Note that the
CO$_2$/H$_2$O ratios toward B5 IRS1 and HH46 IRS are significantly
larger compared to the average over many, mostly massive protostellar
sightlines (0.17$\pm$0.03; Gerakines et al. 1999).  

\section{Discussion}~\label{sec:disc}

\subsection{Evolution of Ices in Low Mass Environments}~\label{sec:disc1}

The formation and evolution of interstellar ices is, in principle,
strongly dependent on local conditions such as the atomic hydrogen and
carbon density, the temperature, the cosmic ray flux, the ultraviolet
photon flux, and the presence of shocks. Thus, key issues are the
location of the ices along the absorption line of sight and the
relative contributions from foreground clouds, envelopes, and inclined
disks. The continuous rise of both the B5 IRS1 and HH46 IRS SEDs
between 3 and 40 \mum, as well as the detection of extended
submillimeter emission in JCMT archive images, are in favor of
envelope-dominated models. We established the properties of these
envelopes using the approach of J{\o}rgensen et al. (2002). Adopting
optical constants for bare and ice coated silicate grains (Ossenkopf
\& Henning 1994) the 2-2000 \mum\ SED and the depth of the observed
superposed ice and silicate bands were self-consistently modeled.  The
SED and silicate band depth of B5 IRS1 are well fitted by the envelope
models, and a significant fraction of the ices has evaporated. In
contrast to B5 IRS1, the submillimeter/far-infrared and mid-infrared
SST SED of HH46 IRS cannot be simultaneously fitted. Possibly, this
envelope is embedded in a larger scale, cold cloud not modeled within
the framework of the simple spherical envelope.

\begin{table}[t]
\center
\caption{Column densities and Abundances}~\label{t1}
\begin{tabular}{lcccc}
\noalign{\smallskip}
\hline
\noalign{\smallskip}
Quantity                            & B5 IRS1   & HH46   & N7538/9$^{\rm e}$ & W33A$^{\rm f}$   \\
\noalign{\smallskip}
\hline
\noalign{\smallskip}
$N$(CO$_2$)/$N$(H$_2$O)             & 0.37      & 0.32   &  0.24       & 0.16   \\
$N$(CO-p)/$N$(H$_2$O)$^{\rm a}$     & 0.16      & 0.15   &  0.02       & 0.08   \\
$N$(CO-np)/$N$(H$_2$O)$^{\rm b}$    & 0.27      & 0.05   &  0.14       & 0.03   \\
$N$(CH$_3$OH)/$N$(H$_2$O)           & $<$0.06   & 0.07   &  0.07       & 0.22   \\
$N$(OCN$^-$)/$N$(H$_2$O)$^{\rm c}$  & $<0.005$  & 0.007  &  0.006      & 0.02   \\
\noalign{\smallskip}
\hline
\noalign{\smallskip}
$N$(H$_2$O) [10$^{18}$ \sqcm]       & 2.2 (0.3) & 8.0    &  6.8        & 9.0    \\
$N$(H$_2$O)/$N_{\rm H}$ [10$^{-5}$] & 5.2       & 5.7    &  5.0        & 3.2    \\
$N_{\rm H}$ [10$^{22}$\sqcm]        & 4$^{\rm d}$& 14$^{\rm d}$&  16   & 28     \\
\noalign{\smallskip}
\hline
\noalign{\smallskip}
\multicolumn{5}{p{8cm}}{$^{\rm a}$ broad `polar' CO component, likely H$_2$O mixture}\\
\multicolumn{5}{p{8cm}}{$^{\rm b}$ broad `apolar' CO component, likely pure CO}\\
\multicolumn{5}{p{8cm}}{$^{\rm c}$ using OCN$^-$ band strength of van
 Broekhuizen et al. 2004}\\
\multicolumn{5}{p{8cm}}{$^{\rm d}$ from the envelope models described in the text}\\
\multicolumn{5}{p{8cm}}{$^{\rm e}$ \citealt{whi96} and references therein}\\
\multicolumn{5}{p{8cm}}{$^{\rm f}$ \citealt{gib00} and references therein}\\
\end{tabular}
\end{table}

Next we address the extent to which the ices in these envelopes have
been processed. Several indicators are available. Laboratory
experiments have shown that heating of ice mixtures with
concentrations CO$_2$/H$_2$O$\geq 1$ results in crystallization and an
effective segregation of the CO$_2$ and H$_2$O
species. Spectroscopically this is recognized as double peaked
profiles, characteristic of the pure CO$_2$ matrix (Ehrenfreund et
al. 1998). Depending on the ice composition, the amorphous to
crystalline phase transition occurs at 50-90 K in space, lower than
the corresponding laboratory temperatures due to the longer
interstellar time scales (Boogert et al. 2000).  Substructures are
seen in the HH46 IRS CO$_2$ band (Fig. 2b), but are much weaker
compared to some highly heated massive protostellar envelopes
(Gerakines et al. 1999), and are absent in B5 IRS1. In fact, the
CO$_2$ band profile of HH46 IRS is comparable to that observed toward
one of the least processed envelopes, surrounding the massive
protostar NGC 7538 IRS9.  In the simplest scenario of a single ice at
one temperature, the mixture CH$_3$OH:H$_2$O:CO$_2$=0.3:1:1 (\S 4.2)
must still have a laboratory temperature as high as 115 K, or $\sim$75
K in space (Fig. 2b).  In the more likely scenario of a temperature
gradient along the line of sight, the bulk of the HH46 IRS and B5 IRS1
envelopes have temperatures well below 50 K.  A fraction of the inner
envelope of HH46 IRS must be warmer, causing the observed weak
substructures. Such two component fits explain the observed profiles
well (Fig. 2c).

More extensive processing at lower temperatures most likely has
occurred within the envelopes, however. Unlike the massive protostar
NGC 7538 IRS9, the ground based 4.67 \mum\ spectra of solid CO toward
HH46 IRS and B5 IRS1 show a weak central narrow component (Fig. 2d;
Table 1).  The most volatile CO--rich `apolar' ices may thus have
evaporated (Tielens et al. 1991). HH46 IRS shows a particularly broad
profile, resembling the massive protostar W33A. Like W33A, HH46 IRS
shows an absorption feature at 4.62 \mum, likely attributed to the
OCN$^-$ species. This molecule may be produced from HNCO in the solid
state at relatively low ice temperatures $<$50 K (van Broekhuizen et
al. 2004).

Concluding, while high temperature ice processing, traced by the
CO$_2$ band, is not observed in the low mass envelopes, low
temperature ($<50$ K) processing may play a significant
role. Qualitatively this is similar to some high mass protostars (W33
A, NGC7538 IRS9). The proposed evolutionary sequence of ice processing
in massive envelopes (Boogert et al. 2000; van der Tak et al. 2000) is
indeed largely based on high temperature indicators, such as ice
crystallization, hot core gas temperatures and gas/solid state ratios.
These relations need to be investigated in a larger sample of low mass
protostars, in which the CO$_2$ band profile is a crucial tracer.

\subsection{CH$_3$OH:CO$_2$ Complexes}~\label{sec:disc2}

The long wavelength wing of the CO$_2$ bending mode may be a tracer of
the presence of CH$_3$OH in the ices.  CO$_2$ and CH$_3$OH form
complexes, leading to an enhanced wing in some sources well fitted by
laboratory ices with a CH$_3$OH:H$_2$O:CO$_2$ mixing ratio of 1:1:1
(e.g. W33A in Fig. 2b). Recently, large abundances of CH$_3$OH of 25\%
with respect to H$_2$O were found in the envelopes of some low mass
protostars (Pontoppidan et al. 2003). Clearly, B5 IRS1 and HH46 IRS
both have lower CH$_3$OH abundances (Table 1). This is consistent with
the weakness of the long wavelength CO$_2$ wing.  Indeed, laboratory
mixtures with a CH$_3$OH:H$_2$O:CO$_2$ ratio of 0.3:1:1 fit the
observed band well. The high CO$_2$/CH$_3$OH column density ratio of
$\sim$6 toward these sources is however barely consistent with this
laboratory mixture.  Alternatively, the band profile can be explained
by the combination of an abundant H$_2$O--rich ice responsible for the
long wavelength wing, and an at least partly heated CO$_2$-rich ice
responsible for the crystallization substructures seen in HH46 IRS
(Fig. 2c).

\section{Conclusions and Future Work}~\label{sec:concl}

High quality SST observations of the CO$_2$ bending mode at 15 \mum\
toward low mass protostars offer a new tracer of ice mantle
composition and evolution.  The embedded low mass systems B5 IRS1 and
HH46 IRS show CH$_3$OH-poor ices with little evidence for 50-90 K
thermal processing in their envelopes.  Lower temperature processing
appears evident in the solid CO band. These results form the basis for
future studies on the physical and chemical state of ices entering
protoplanetary disks, and how these and solar system ices are related.
CO$_2$ bending mode observations of more evolved systems and edge-on
disks are required.

\acknowledgments

Support for this work, part of the Spitzer Space Telescope Legacy
Science Program, was provided by NASA through Contract Numbers 1224608
and 1230780 issued by the Jet Propulsion Laboratory, California
Institute of Technology under NASA contract 1407.  Astrochemistry in
Leiden is supported by a NWO Spinoza grant and a NOVA grant.


\end{document}